\documentclass{article}
\usepackage{amsmath,graphicx,mlspconf}
\usepackage{xcolor}
\usepackage{booktabs} 
\usepackage[hidelinks]{hyperref}
\usepackage{amsfonts}

\copyrightnotice{%
\parbox{0.94\textwidth}{%
\fontsize{5.5}{6.2}\selectfont
{\copyright} 2026 IEEE. Personal use of this material is permitted. Permission from IEEE must be obtained for all other uses, in any current or future media, including reprinting/republishing this material for advertising or promotional purposes, creating new collective works, for resale or redistribution to servers or lists, or reuse of any copyrighted component of this work in other works.
}}

\toappear{2026 IEEE International Workshop on Machine Learning for Signal Processing, Sep.\ 28-- Oct.\ 1, 2026, Atlanta, USA}

\title{BAT-CLIP: Trimodal Alignment of Brain, Audio and Text}
\name{%
    \parbox{0.95\textwidth}{%
    \centering
    \itshape
    Suhyun Kim$^{1,*}$,
    Jinmo Han$^{2,*}$,
    Danny Dongyeop Han$^{2,*}$,
    Ahhyun Lucy Lee$^{2}$,
    Jewoon Lee$^{2}$,
    Yonghyeon Gwon$^{2}$,
    Zach Paris$^{3}$,
    Chun Kee Chung$^{2}$,
    Saewoong Bahk$^{2}$,
    Nam Soo Kim$^{2}$,
    Seong~Jae~Hwang$^{1}$,
    Jiook Cha$^{2,\dagger}$%
    \thanks{$^{*}$These authors contributed equally to this work.
            $^{\dagger}$Corresponding author: connectome@snu.ac.kr.}%
    }%
}
\address{%
    $^{1}$ Yonsei University, Seoul, Republic of Korea \\
    $^{2}$ Seoul National University, Seoul, Republic of Korea \\
    $^{3}$ Dartmouth College, Hanover, NH, USA%
}

\begin{document}

\maketitle

\begin{abstract}
Decoding and interpreting naturalistic speech from the brain increasingly relies on alignment to pretrained speech and language representation spaces. However, current CLIP-style brain--speech alignment ground neural activity to a single anchor modality---audio or text---despite the brain’s inherently multimodal speech processing. This induces a trade-off: audio anchoring preserves temporal structure but weakens linguistic separability, while text anchoring captures semantics yet discards acoustic detail.
We propose \textbf{\emph{BAT-CLIP}}, the first CLIP-style trimodal alignment framework for iEEG that jointly aligns neural embeddings to both pretrained audio and text anchors in a shared, frozen audio--text manifold. On the naturalistic Podcast benchmark, \emph{BAT-CLIP} yields more robust representations than bimodal CLIP baselines. We also highlight the importance of using self-supervised foundation models for CLIP training.
\end{abstract}

\newcommand{\cem}[1]{\textcolor{blue}{cem: #1}}

\section{Introduction}

Brain--speech alignment is a central objective in neurolinguistics, spanning \emph{encoding} models that predict neural responses from speech/language representations \cite{goldstein25-unified} and \emph{decoding} models that infer speech or linguistic variables from brain activity \cite{Defossez23-NatMI, Mishra25-Thought2Text}.  
It matters for applications because stronger alignment improves downstream decoding, enabling simple readouts to recover linguistic variables and more expressive decoders to generate coherent text. 
It is also a scientific target, offering a principled way to test which computational representations best match the brain and where different representational levels emerge across cortex \cite{Wang25-LingNeuralDecoding,jain24-computational}.  

Recently, large-scale speech and language foundation models have reshaped brain--speech alignment: rather than relying on hand-crafted features or task-specific labels, neural signals can be aligned end-to-end to pretrained embedding spaces via CLIP-style contrastive objectives \cite{Yang24-NeuSpeech, Liu26-MindMix}. This approach has driven strong gains in both encoding and decoding---including naturalistic speech and vision in fMRI and electrophysiology (EEG/MEG/iEEG)---by leveraging the semantic structure and inductive biases of pretrained audio--text models \cite{Wang25-LingNeuralDecoding}. 

\begin{figure}[t!]
  \centering
  \includegraphics[width=0.96\linewidth]{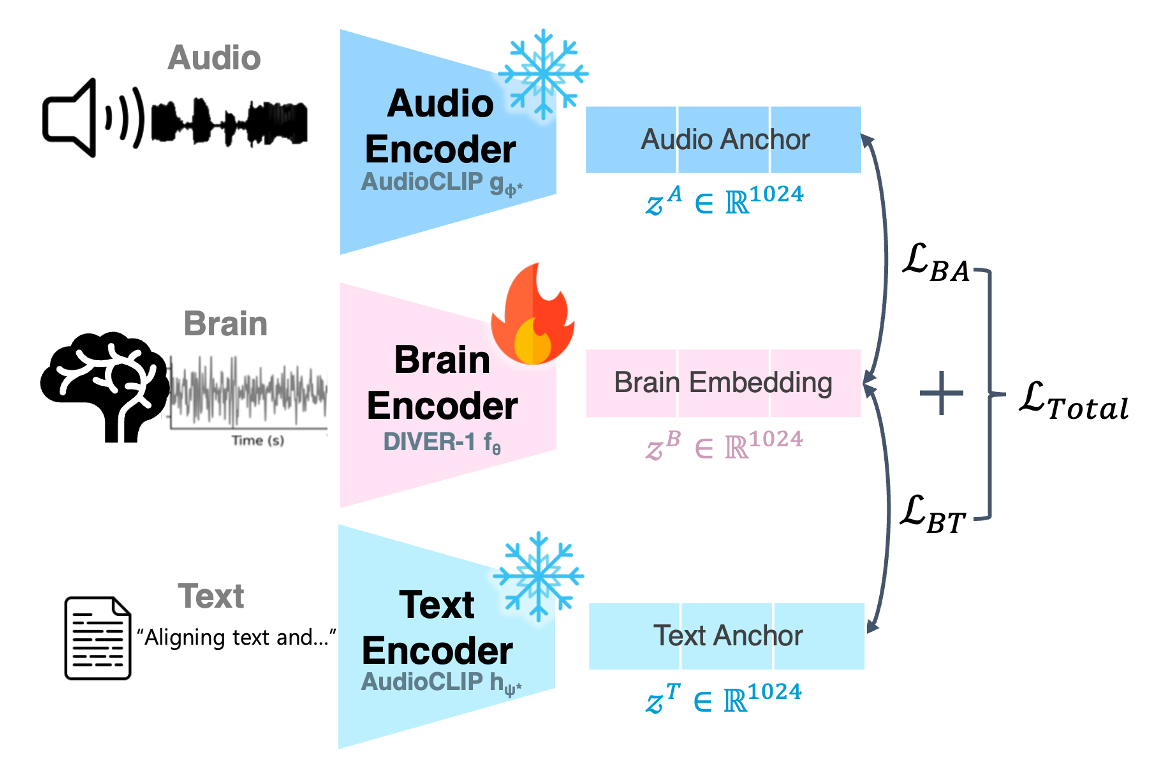}
  \caption{\textbf{Overview of \textit{BAT-CLIP}}. A trainable brain encoder $f_\theta$, initialized from brain foundation model DIVER-1, maps neural windows to brain embeddings $z^B$. Frozen \mbox{AudioCLIP} encoders provide audio anchors $z^A$ and text anchors $z^T$. \textit{\mbox{BAT-CLIP}} optimizes a trimodal contrastive objective $\mathcal{L}_{Total}$ to align neural embeddings with the joint audio–text representation space.}
  \label{fig:framework}
\end{figure}

However, current CLIP-style brain--speech methods anchor neural activity to a single modality (audio or text). This neglects the multimodal nature of speech comprehension, where the brain integrates linguistic and acoustic information across distributed networks \cite{McGettigan12-Neuropsychologia}. Encoding work further shows that audio/speech and text/language models capture complementary aspects of neural language comprehension: speech models better explain activity in earlier auditory regions, whereas text models better explain later language regions linked to higher-level linguistic processing \cite{Oota24-SpeechLMs,antonello23-scaling}. 


Motivated by this complementarity of audio and text model representations, we propose \emph{BAT-CLIP} (Brain--Audio--Text CLIP), which aligns a pretrained brain foundation model to a frozen audio--text manifold (AudioCLIP \cite{Guzhov22-AudioCLIP}) via a CLIP-style \emph{trimodal} contrastive objective, anchoring brain representations to both audio and text embeddings (Fig.~\ref{fig:framework}). Furthermore, we assess the value of self-supervised pretraining itself by comparing random initialization to a strong SSL-pretrained brain foundation model, showing that SSL provides a critical prior for reliable trimodal alignment. We further perform embedding-space analyses that characterize how trimodal alignment better organizes neural representations relative to audio and text anchors.

\section{Methods}
\label{sec:methods}

We propose \textbf{\emph{BAT-CLIP}}, a trimodal contrastive alignment framework that grounds self-supervised pretrained brain representations in a pretrained audio–text manifold. Specifically, we initialize from a brain foundation model (DIVER-1 \cite{Han25-DIVER}) and adapt with a trimodal objective using paired iEEG–audio–text data.

\subsection{Problem Formulation} 

Let \(\mathcal{D}=\{(x_i^{B},x_i^{A},x_i^{T})\}_{i=1}^{N}\) denote time-aligned triplets, where \(x_i^{B}\) is an iEEG segment with \(C\) channels over \(T\) time points, and \(x_i^{A}\) and \(x_i^{T}\) are the corresponding audio and transcript segments.
We downsample iEEG to match the DIVER-1 checkpoint, resample audio to 44.1\,kHz for AudioCLIP, and tokenize transcripts with the AudioCLIP tokenizer.

Our CLIP alignment objective is to train a brain encoder $f_\theta: \mathcal{X}^B \to \mathbb{R}^{d}$ that maps neural windows $x^B_i$ to a brain embedding $z^B_i = f_\theta(x^B_i)$ in a joint embedding space ($d=1024$) shared with audio and text representations.

\subsection{Alignment Prior: DIVER-1 Initialization}
\label{methods:alignment_prior}

We initialize the brain encoder from DIVER-1, a iEEG foundation model pretrained with self-supervised masked reconstruction on large amounts of data. This initialization acts as an alignment prior, reducing overfitting in our limited paired brain--audio--text dataset. During CLIP alignment, we flatten the brain encoder's embeddings and project them to the shared embedding space with a lightweight two-layer MLP, followed by \(\ell_2\) normalization.

\subsection{Shared Audio-Linguistic Space: AudioCLIP Anchors}
\label{methods:anchors}

We ground neural representations in a shared multimodal space using the pretrained AudioCLIP audio and text encoders. For each time-aligned triplet, we encode the audio waveform and transcript with \(g_{\phi^*}\) and \(h_{\psi^*}\) to obtain anchor embeddings
\(z_i^{A}=\mathrm{norm}(g_{\phi^*}(x_i^{A}))\) and \(z_i^{T}=\mathrm{norm}(h_{\psi^*}(x_i^{T}))\),
where \(\mathrm{norm}(\cdot)\) denotes \(\ell_2\) normalization. We keep \(\phi^*\) and \(\psi^*\) frozen throughout training so that the brain encoder is trained to match a stable, well-regularized audio--text manifold rather than co-adapting the anchor spaces in the low-data setting.

\subsection{Trimodal Contrastive Alignment}
\label{methods:alignment}

We align brain embeddings to both audio and text anchors with a symmetric \textit{Brain--Text and Brain--Audio} contrastive objective. Over the global batch of size \(N_b\), we compute similarity logit matrices \(\mathbf{S}_{BA}, \mathbf{S}_{BT}\in\mathbb{R}^{N_b\times N_b}\): $\mathbf{S}_{BA} = \alpha\, \mathbf{Z}_B \mathbf{Z}_A^\top, \ \mathbf{S}_{BT} = \alpha\, \mathbf{Z}_B \mathbf{Z}_T^\top,$ where $\mathbf{Z}_B, \mathbf{Z}_A, \mathbf{Z}_T \in \mathbb{R}^{N_b \times d}$ stack the $\ell_2$-normalized embeddings $z^B_i, z^A_i, z^T_i$ for the batch. The constant $\alpha = 14.3$ is a scaling factor applied to sharpen the logit distribution and stabilize optimization.

Following CLIP, we compute a bidirectional cross-entropy loss for each pair. Given diagonal ground-truth labels $\mathbf{y} = [0,1,\dots,N_b-1]$, the Brain-Audio loss $\mathcal{L}_{BA}$ and the Brain-Text loss $\mathcal{L}_{BT}$ are formulated as:
\begin{align}
    \mathcal{L}_{BA} &= \frac{1}{2}\left( \mathrm{CE}(\mathbf{S}_{BA}, \mathbf{y}) + \mathrm{CE}(\mathbf{S}_{BA}^\top, \mathbf{y}) \right), \\
    \mathcal{L}_{BT} &= \frac{1}{2}\left( \mathrm{CE}(\mathbf{S}_{BT}, \mathbf{y}) + \mathrm{CE}(\mathbf{S}_{BT}^\top, \mathbf{y}) \right).
\end{align}

The final contrastive objective is the sum:
\begin{equation}
    \mathcal{L}_{Total} =  \frac{1}{2}\left(\mathcal{L}_{BA} + \mathcal{L}_{BT} \right)
\end{equation}

\begin{table*}[t]
\centering
\footnotesize
\renewcommand{\arraystretch}{1.10}
\setlength{\tabcolsep}{3.0pt}


\begin{tabular}{
l c@{\hspace{6pt}}
cc@{\hspace{6pt}} cc@{\hspace{6pt}} cc@{\hspace{6pt}} cc@{\hspace{6pt}} cc@{\hspace{6pt}} cc
}
\toprule
\textbf{Method} & \textbf{Type} &
\multicolumn{2}{c}{\textbf{Content}} & 
\multicolumn{2}{c}{\textbf{Onset}} & 
\multicolumn{2}{c}{\textbf{POS}} & 
\multicolumn{2}{c}{\textbf{GPT Surprisal}} & 
\multicolumn{2}{c}{\textbf{Word Emb}} & 
\multicolumn{2}{c}{\textbf{Whisper}} \\
\cmidrule(lr){3-4}\cmidrule(lr){5-6}\cmidrule(lr){7-8}\cmidrule(lr){9-10}\cmidrule(lr){11-12}\cmidrule(lr){13-14}
& &
\textbf{AUC} & \textbf{$\Delta$} & 
\textbf{AUC} & \textbf{$\Delta$} & 
\textbf{AUC} & \textbf{$\Delta$} & 
\textbf{AUC} & \textbf{$\Delta$} & 
\textbf{AvgAUC} & \textbf{$\Delta$} & 
\textbf{Acc} & \textbf{$\Delta$} \\
\midrule
CNN(Baseline)$^{\dagger}$ & Sup &
0.5498 & \multicolumn{1}{c}{--} & 
\underline{0.7519} & \multicolumn{1}{c}{--} & 
0.5653 & \multicolumn{1}{c}{--} & 
0.5019 & \multicolumn{1}{c}{--} & 
0.6260 & \multicolumn{1}{c}{--} & 
0.6634 & \multicolumn{1}{c}{--} \\
\midrule
BrainBERT & SSL & 0.5505 & +0.13\% & 0.6341 & -15.67\% & 0.5325 & -5.80\% & 0.5157 & +2.75\% & 0.5533 & -11.61\% & 0.5805 & -12.49\% \\
PopT      & SSL & 0.4950 & -9.97\% & 0.4654 & -38.10\% & 0.5031 & -11.00\% & 0.5022 & +0.06\% & 0.5002 & -20.10\% & 0.5014 & -24.42\% \\
DIVER-1   & SSL & 0.5645 & +2.67\% & \textbf{0.7729} & \textbf{+2.79\%} & 0.5656 & +0.05\% & 0.5155 & +2.71\% & 0.6450 & +3.04\% & 0.7022 & +5.85\% \\
\midrule
\emph{BA-CLIP}   & Align & \textbf{0.5840} & \textbf{+6.31\%} & 0.6560 & -12.61\% & 0.5673 & +0.32\% & 0.5169 & +3.02\% & 0.6468 & +4.26\% & 0.7141 & +7.75\% \\
\emph{BT-CLIP}   & Align & 0.5813 & +5.82\% & 0.6562 & -12.56\% & \underline{0.5709} & \underline{+0.97\%} & \underline{0.5203} & \underline{+3.69\%} & \underline{0.6510} & \underline{+4.94\%} & \underline{0.7157} & \underline{+7.98\%} \\
\midrule
\textbf{\emph{BAT-CLIP(Ours)}} & Align &
\underline{0.5819} & \underline{+5.96\%} & 0.6589 & -12.18\% & \textbf{0.5740} & \textbf{+1.50\%} & \textbf{0.5229} & \textbf{+4.21\%} & \textbf{0.6512} & \textbf{+5.01\%} & \textbf{0.7162} & \textbf{+8.09\%} \\
\bottomrule
\end{tabular}

\caption{\textbf{Overall decoding performance} (averaged over 9 subjects). We report AUC for \textit{Content}, \textit{Onset}, \textit{POS}, \textit{GPT surprisal}, and \textit{Word Emb} decoding, and accuracy for \textit{Whisper}. $\Delta$ denotes relative gain over the supervised CNN baseline (baseline $\Delta$: ``--''). $\dagger$ indicates full fine-tuning (CNN); all other methods are evaluated via linear probing. \textit{BA-CLIP}, \textit{BT-CLIP}, and \textit{\mbox{BAT-CLIP}} apply CLIP-style contrastive alignment on top of the pretrained DIVER-1 brain encoder, using brain--audio, brain--text, and trimodal brain--audio--text alignment, respectively.  Overall, DIVER-1 is the strongest foundation-model baseline, CLIP alignment generally improves over DIVER-1, and BAT-CLIP achieves the best result among the aligned variants on five of six probe tasks. \textbf{Bold} and \underline{underline} denote the best and second-best results within each column.}
\label{tab:g1_thin}
\end{table*}

\section{Experiments and Results}
\label{sec:experiments}

We evaluate \emph{BAT-CLIP} with the goal of isolating the effect of \emph{trimodal alignment} on iEEG representations under realistic low-data regimes. We first describe the experimental setup---data, preprocessing, baselines, and training configurations for alignment learning. We then report quantitative results on the Podcast Benchmark probe suite and analyze ablations that attribute performance to specific design choices. 

\subsection{Experimental Settings}
\label{subsec:setup}

\textbf{Data and benchmark.}
We use the \emph{Podcast} intracranial electrophysiology dataset\cite{zada25-podcast}, which provides iEEG recordings during naturalistic passive listening to a spoken narrative, along with word-level time-aligned transcripts and derived linguistic/acoustic features. All reported results follow the \emph{Podcast Benchmark}
\footnote{Paris, Z., Sim, S., Bhattacharjee, A., Jalon, I., Peeper, G., Han, J., \& Hasson, U. \emph{ECoG Podcast Benchmark} [Computer software]. Available at \url{https://github.com/hassonlab/podcast-benchmark}.}
evaluation suite built on top of this dataset, which standardizes probe tasks, metrics, and data splits for comparable decoding-based assessment.

\noindent\textbf{Baselines and model variants.}
We compare against (i) a supervised CNN baseline trained end-to-end with full finetuning on each downstream probe task and (ii) recent brain foundation models (BrainBERT\cite{wang23-brainbert}, PopT\cite{chau25-popt}, and DIVER-1\cite{Han25-DIVER}). Since \emph{unaligned} DIVER-1 provides the strongest off-the-shelf performance in this setting, we adopt its pretrained 12-layer transformer backbone ($d_{\text{model}}=256$) as the common initialization for our aligned variants (\emph{BA-CLIP} (Brain-Audio CLIP), \emph{BT-CLIP} (Brain-Text CLIP), and \emph{BAT-CLIP} (Brain-Audio-Text CLIP)). This controls for architecture and initialization, attributing gains to multimodal alignment.

\noindent\textbf{Preprocessing.}
For each subject, we exclude bad channels and retain the top 10\% most statistically responsive electrodes (selected using training data only to avoid leakage). Raw iEEG is standardized and resampled to 500\,Hz, then segmented into 1.0\,s windows anchored to benchmark-defined target events. Following DIVER-1, each window is tokenized into non-overlapping patches of 50 samples (0.1\,s) to form the transformer input sequence.

\noindent\textbf{Alignment.}
Aligned models are trained with the contrastive objectives described in Sec.~\ref{sec:methods}, using frozen audio/text anchor encoders. We optimize the brain encoder and its projection head; the pretrained anchor encoders remain fixed throughout training. Hyperparameters for alignment training (e.g., learning rate and weight decay) are selected on the validation split and held fixed when reporting test performance.

\noindent\textbf{Evaluation tasks, metrics and protocol.}
We evaluate each model on six probe tasks that test what information can be decoded from neural activity around word/sentence events: (1) content vs.\ non-content word classification (binary), (2) coarse part-of-speech classification (5 classes), (3) sentence-onset detection (binary; negatives are sampled away from true onsets within the same sentence), (4) GPT Surprisal (GPT-2 XL; 3-bin: low/medium/high), (5) word-embedding decoding (GPT-2 XL/GloVe), and (6) Whisper latent decoding\cite{goldstein25-unified}. For classification probes (1)–(4), we report AUC (binary ROC-AUC or one-vs-rest multiclass AUC). For word-embedding decoding, we report AvgAUC: one-vs-rest AUC is computed per vocabulary item and then averaged across eligible words. For Whisper-latent decoding, we report pairwise matching accuracy: for each predicted segment, we test whether it is more similar to its true matched target than to mismatched targets. We train supervised head for each task under an identical finetuning protocol: a fixed temporal 60\%/20\%/20\% train/validation/test split over five contiguous blocks, with the first block held out for testing and the remaining blocks split sequentially into training and validation, lag fixed at $0$ ms, and early stopping with a maximum of 30 epochs and patience $=5$.

\begin{table*}[t]
\centering
\footnotesize
\renewcommand{\arraystretch}{1.05}
\begin{tabular}{
ll
cc@{\hspace{6pt}} cc@{\hspace{6pt}} cc@{\hspace{6pt}} cc@{\hspace{6pt}} cc@{\hspace{6pt}} cc
}
\toprule
\textbf{Method} & \textbf{Init.} & 
\multicolumn{2}{c}{\textbf{Content}} & 
\multicolumn{2}{c}{\textbf{Onset}} & 
\multicolumn{2}{c}{\textbf{POS}} & 
\multicolumn{2}{c}{\textbf{GPT Surprisal}} & 
\multicolumn{2}{c}{\textbf{Word Emb}} & 
\multicolumn{2}{c}{\textbf{Whisper}} \\
\cmidrule(lr){3-4}\cmidrule(lr){5-6}\cmidrule(lr){7-8}\cmidrule(lr){9-10}\cmidrule(lr){11-12}\cmidrule(lr){13-14}
& & 
\textbf{AUC} & \textbf{$\Delta$} & 
\textbf{AUC} & \textbf{$\Delta$} & 
\textbf{AUC} & \textbf{$\Delta$} & 
\textbf{AUC} & \textbf{$\Delta$} & 
\textbf{AvgAUC} & \textbf{$\Delta$} & 
\textbf{Acc} & \textbf{$\Delta$} \\
\midrule
\textbf{\emph{\emph{\emph{BA-CLIP}}}} & Scratch & 0.5671 & +3.14\% & \textbf{0.7792} & \textbf{+3.75\%} & 0.5624 & -0.35\% & 0.5143 & +2.55\% & 0.6059 & -2.33\% & 0.6609 & +0.05\% \\
 & DIVER-1 & \textbf{0.5840} & \textbf{+6.31\%} & 0.6560 & -12.61\% & \textbf{0.5673} & \textbf{+0.32\%} & \textbf{0.5169} & \textbf{+3.02\%} & \textbf{0.6468} & \textbf{+4.26\%} & \textbf{0.7141} & \textbf{+7.75\%} \\
\midrule
\textbf{\emph{BT-CLIP}} & Scratch & 0.5650 & +2.85\% & \textbf{0.7822} & \textbf{+4.19\%} & 0.5635 & -0.18\% & 0.5120 & +2.10\% & 0.6296 & +1.52\% & 0.6607 & -0.02\% \\
 & DIVER-1 & \textbf{0.5813} & \textbf{+5.82\%} & 0.6562 & -12.56\% & \textbf{0.5709} & \textbf{+0.97\%} & \textbf{0.5203} & \textbf{+3.69\%} & \textbf{0.6510} & \textbf{+4.94\%} & \textbf{0.7157} & \textbf{+7.98\%} \\
\midrule
\textbf{\emph{BAT-CLIP}} & Scratch & 0.5586 & +1.70\% & \textbf{0.7600} & \textbf{+0.94\%} & 0.5480 & -2.94\% & 0.5119 & +2.09\% & 0.6123 & -1.32\% & 0.6463 & -2.36\% \\
 & DIVER-1 & \textbf{0.5819} & \textbf{+5.96\%} & 0.6589 & -12.18\% & \textbf{0.5740} & \textbf{+1.50\%} & \textbf{0.5229} & \textbf{+4.21\%} & \textbf{0.6512} & \textbf{+5.01\%} & \textbf{0.7162} & \textbf{+8.09\%} \\
\bottomrule
\end{tabular}
\caption{\textbf{Effect of pretraining across methods.} We compare CLIP-aligned models trained from scratch (Scratch) versus the same models initialized with pretrained DIVER-1 weights, across unimodal alignments (\emph{BA-CLIP}: brain--audio; \emph{BT-CLIP}: brain--text) and trimodal alignment (\emph{BAT-CLIP}: brain--audio--text). As in Table~\ref{tab:g1_thin}, $\Delta$ denotes relative gain over the supervised CNN baseline (baseline $\Delta$: ``--''). Pretrained DIVER-1 initialization is key to realizing the gains from trimodal CLIP alignment; training from scratch largely fails to capitalize on brain--audio--text supervision.}
\label{tab:ablation_init}
\end{table*}

\begin{figure*}[t]
  \centering
  \includegraphics[width=\textwidth]{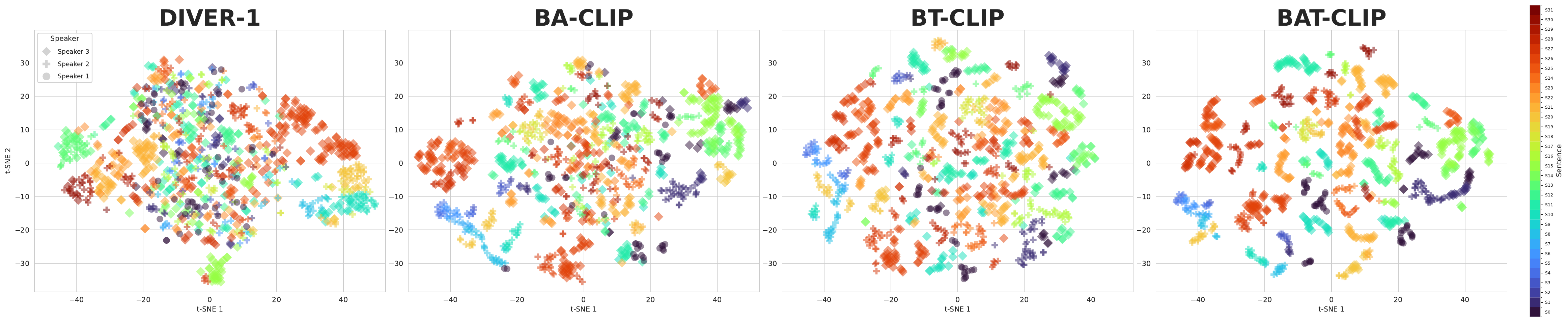}
  \caption{\textbf{Visualization of learned word-level representations.} 2D t-SNE projection of brain embeddings extracted by various models for Sub03. Each point represents a single word-level embedding. The color of the point denotes the sentence to which the word belongs, and the shape indicates the speaker. Compared to the baselines, the proposed trimodal alignment model, \textit{BAT-CLIP}, exhibits tighter clustering of words from the same sentence (shared colors), consistent with the sentence-level clustering results in Table~\ref{tab:clustering}.}
  \label{fig:tsne}
\end{figure*}

\vspace{-1pt}
\subsection{Benefits of Trimodal Alignment.}

Table~\ref{tab:g1_thin} compares our \emph{BAT-CLIP} against bimodal alignment variants built on the same DIVER-1 brain encoder---\emph{BA-CLIP} (Brain--Audio) and \emph{BT-CLIP} (Brain--Text)---as well as a supervised CNN baseline and pretrained foundation models (BrainBERT, PopT, DIVER-1). Among the foundation-model and CNN baselines, DIVER-1 is the strongest overall, despite being evaluated via linear probing rather than full fine-tuning; this motivates using DIVER-1 pretrained weights as the initialization for CLIP-style alignment.

Applying CLIP alignment on top of DIVER-1 generally improves decoding performance, and \emph{BAT-CLIP} is the best aligned model overall. Trimodal alignment outperforms the bimodal variants on POS, GPT surprisal, word-embedding decoding, and Whisper decoding; on the Content probe, however, it falls behind the bimodal \emph{BA-CLIP}. A notable exception is the Onset task, where both the CNN and unaligned DIVER-1 baselines remain strongest and CLIP alignment degrades performance. We hypothesize this reflects a mismatch between the objective and the task structure. 
Onset detection is timing-critical and can depend on transient, highly localized neural signatures (sometimes confined to a small subset of channels) \cite{Yulia19-SCIADV}, whereas CLIP alignment encourages temporal invariances and relies on globally pooled window-level embeddings. Such pooling can wash out sharp transient cues essential for precise onset detection.

\subsection{SSL Initialization for Trimodal Alignment.}

Table~\ref{tab:ablation_init} compares CLIP-aligned models trained from scratch (Random Init.) with the same objectives initialized from pretrained DIVER-1 weights. Pretrained initialization substantially amplifies the benefits of CLIP alignment, yielding markedly stronger downstream decoding. This matches the standard recipe in other domains, where CLIP-style objectives are typically applied \emph{on top of} strong self-supervised or foundation-model priors \cite{li22-blip,elizalde23-clap}. In contrast, prior brain CLIP-alignment work has rarely combined SSL pretraining with subsequent CLIP grounding. Taken together, these results position brain foundation models as a strong substrate for multimodal grounding and suggest that scaling and reusing pretrained neural representations will be central to future progress in brain--audio--text alignment, in line with recent results EEG \cite{Liu26-MindMix}.

Notably, the advantage of \emph{trimodal} alignment is only clearly realized when starting from a strong SSL prior: with DIVER-1 initialization, \emph{BAT-CLIP} benefits from joint brain--audio--text supervision, whereas under random initialization the trimodal setting does not reliably separate from unimodal variants, suggesting pretrained representations are needed to effectively absorb and exploit trimodal CLIP supervision.

\subsection{Embedding-space analysis}
\label{sec:embedding_analysis}



{\looseness=-1 We analyzed how SSL pretraining and CLIP-style grounding shape the geometry of neural representations by comparing \textit{test-fold neural embedding spaces} from DIVER-1, \textit{BA-CLIP}, \textit{BT-CLIP}, and \textit{BAT-CLIP}. For each test segment, we extracted the \(\ell_2\)-normalized brain backbone embedding \(z_i^B\) and visualized the space with t-SNE (Fig.~\ref{fig:tsne}). SSL-only DIVER-1 exhibits weak sentence-level structure, with embeddings from different sentences substantially mixed, whereas CLIP clearly re-organizes the space, with \emph{BAT-CLIP} showing the strongest sentence-level grouping.\par}


To quantify this, we evaluate sentence-level clustering of word-aligned neural embeddings using two metrics (Table~\ref{tab:clustering}), over \(K{=}31\) test sentences with at least three words (\(N{=}1{,}025\)). \textbf{Cosine Coherence} is computed as \(\mathrm{CC}=\frac{1}{K}\sum_s[\overline{\cos}_{\mathrm{intra}}(s)-\overline{\cos}_{\mathrm{inter}}(s)]\), where the terms denote mean cosine similarity among embeddings within sentence \(s\) and between embeddings in \(s\) and those in all other sentences, respectively. \textbf{Separation Ratio} is computed in PCA-reduced 50-dimensional space as \(\mathrm{SR}=\bar{D}_{\mathrm{inter}}/\bar{D}_{\mathrm{intra}}\), where \(\bar{D}_{\mathrm{inter}}\) is the mean distance between sentence centroids and \(\bar{D}_{\mathrm{intra}}\) is the mean within-sentence spread. \emph{BAT-CLIP} attains the highest scores on both metrics, consistent with tighter intra-sentence grouping and stronger inter-sentence separation than DIVER-1 and the bimodal variants.

\begin{table}[t]
\centering
\footnotesize
\renewcommand{\arraystretch}{1.1}
\setlength{\tabcolsep}{5pt}

\begin{tabular}{lcc}
\toprule
\textbf{Method} & \textbf{Cosine Coherence ($\uparrow$)} & \textbf{Separation Ratio ($\uparrow$)} \\
\midrule
DIVER-1 (SSL) & 0.0004 & 0.95 \\
\emph{\emph{\emph{BA-CLIP}}} & 0.0789 & 1.09 \\
\emph{BT-CLIP} & 0.0954 & 1.04 \\
\textbf{\emph{BAT-CLIP} (Ours)} & \textbf{0.1572} & \textbf{1.23} \\
\bottomrule
\end{tabular}
\caption{\textbf{Quantitative clustering evaluation of word-level embeddings.} We evaluate the structure of the learned embedding space for a representative subject (Sub03; Fig.~\ref{fig:tsne}) using sentence-level clustering metrics, where higher values indicate tighter grouping of word embeddings within the same sentence. The proposed trimodal alignment (\textit{BAT-CLIP}) yields the strongest clustering compared to the unaligned foundation model (DIVER-1) and unimodal alternatives.} 
\label{tab:clustering}
\end{table}

\section{Discussion and Conclusion}
\label{sec:discussion}

In this work, we introduce \textbf{\emph{BAT-CLIP}}, a trimodal brain--audio--text alignment framework for learning semantically grounded representations from naturalistic iEEG. By aligning iEEG embeddings to a shared audio--text manifold with contrastive objectives against both audio and text anchors, our results suggest two takeaways: (1) trimodal grounding is effective, yielding stronger performance on linguistic/semantic probes than single-anchor alignment; and (2) SSL-pretrained neural features are key for alignment, as initializing from SSL (DIVER-1) and then applying CLIP-style grounding consistently outperforms training from scratch and enables the most reliable gains from trimodal supervision.

Taken together, these findings motivate a new paradigm for brain--speech alignment: \emph{pretrain neural representations with SSL, then ground them with trimodal contrastive alignment into a shared audio--text manifold}. Concretely, SSL supplies a strong neural prior learned from massive unlabeled recordings (e.g., DIVER-1 pretrained on $\sim$5.3k hours), while the trimodal objective uses both audio and text anchors to inject complementary acoustic-to-semantic structure from only limited paired data (here, $\sim$30 minutes). 


Although we focus on probe tasks to isolate the strengths of the learned representations, this does not directly test end-to-end decoding; a natural next step is to use \textit{BAT-CLIP} as a brain--speech alignment module and attach a downstream LLM decoder to evaluate settings such as open-vocabulary decoding or sequence-level prediction beyond linear probes. In addition, iEEG remains low-resource for naturalistic speech because large datasets are scarce and rarely open, unlike MEG/EEG where larger public corpora are available. How \textit{BAT-CLIP} scales with substantially larger iEEG collections is therefore unclear and may depend on community efforts to curate shared benchmarks, while experiments on large-scale MEG speech datasets would provide a complementary test of cross-modality generalization and scaling.

\section{Acknowledgments}
{\small \sloppy
Work supported by the National Research Foundation of Korea (NRF) grant funded by the Korea government (MSIT) (No. 2021R1C1C1006503, RS-2023-00266787, RS-2023-00265406, RS-2024-00421268, RS-2024-00342301, RS-2024-00435727, RS-2025-25457239, RS-2026-25490183, RS-2026-25524667, RS-2026-25518389, RS-2021-NR061370, NRF-2021M3E5D2A01022515, and NRF-2021S1A3A2A02090597), by the Researchers Program through Seoul National University (No. 200-20250071, 200-20250049, 200-20240057, 0670-20260027, 200-20260009, 200-20250116, 200-20260081, 200-20250115, 200-20250113, 0670-20250039, 200-20260009). Additional support provided by the Institute of Information \& Communications Technology Planning \& Evaluation (IITP) grant funded by the Korea government (MSIT) [No. RS-2021-II211343, Artificial Intelligence Graduate School Program, Seoul National University] and by the Global Research Support Program in the Digital Field (RS-2024-00421268). Also supported by the Artificial Intelligence Industrial Convergence Cluster Development Project funded by the Ministry of Science and ICT and Gwangju Metropolitan City, by the Korea Brain Research Institute (KBRI) basic research program (25-BR-05-01), by the Korea Health Industry Development Institute (KHIDI) and the Ministry of Health and Welfare, Republic of Korea (HR22C1605), and by the Korea Basic Science Institute (National Research Facilities and Equipment Center) grant funded by the Ministry of Education (RS-2024-00435727). We acknowledge the National Supercomputing Center for providing supercomputing resources and technical support (KSC-2023-CRE-0568, KSC-2024-CRE-0198, KSC-2025-CRE-0340).
\par} 

\bibliographystyle{IEEEbib}
\bibliography{strings,refs}

@article{Yulia19-SCIADV,
    author       = {Oganian, Yulia and Chang, Edward F.},
    title        = {A speech envelope landmark for syllable encoding in human superior temporal gyrus},
    journal      = {Science Advances},
    volume       = {5},
    number       = {11},
    pages        = {eaay6279},
    year         = {2019},
    publisher    = {American Association for the Advancement of Science},
}

@article{Defossez23-NatMI,
    author       = {Alexandre D{\'e}fossez and Charlotte Caucheteux and J{\'e}r{\'e}my Rapin and Ori Kabeli and Jean-R{\'e}mi King},
    title        = {Decoding speech perception from non-invasive brain recordings},
    journal      = {Nature Machine Intelligence},
    volume       = {5},
    number       = {10},
    pages        = {1060--1070},
    year         = {2023},
}

@inproceedings{Mishra25-Thought2Text,
    author       = {Abhijit Mishra and Shreya Shukla and Jose Torres and Jacek Gwizdka and Shounak Roychowdhury},
    title        = {Thought2Text: Text Generation from {EEG} Signal using Large Language Models ({LLMs})},
    booktitle    = {Findings of the Association for Computational Linguistics: {NAACL} 2025},
    pages        = {3560--3576},
    year         = {2025},
}

@article{Yang24-NeuSpeech,
    author       = {Yiqian Yang and Yiqun Duan and Qiang Zhang and Hyejeong Jo and Jinni Zhou and Won Hee Lee and Renjing Xu and Hui Xiong},
    title        = {NeuSpeech: Decode Neural signal as Speech},
    journal      = {arXiv preprint},
    volume       = {arXiv:2403.01748},
    year         = {2024},
}

@inproceedings{Liu26-MindMix,
    author       = {Rui Liu and Zhige Chen and Shu Peng and Wenlong You and Zhi-An Huang and Jibin Wu and Kay Chen Tan},
    title        = {MindMix: A Multimodal Foundation Model for Auditory Perception Decoding via Deep Neural-Acoustic Alignment},
    booktitle    = {Proceedings of the International Conference on Learning Representations ({ICLR})},
    year         = {2026},
}

@article{McGettigan12-Neuropsychologia,
    author       = {Carolyn McGettigan and Andrew Faulkner and Irene Altarelli and Jonas Obleser and Harriet Baverstock and Sophie K. Scott},
    title        = {Speech comprehension aided by multiple modalities: Behavioural and neural interactions},
    journal      = {Neuropsychologia},
    volume       = {50},
    number       = {5},
    pages        = {762--776},
    year         = {2012},
}

@inproceedings{Guzhov22-AudioCLIP,
    author       = {Andrey Guzhov and Federico Raue and J{\"o}rn Hees and Andreas Dengel},
    title        = {{AudioCLIP}: Extending {CLIP} to Image, Text and Audio},
    booktitle    = {Proceedings of the {IEEE} International Conference on Acoustics, Speech, and Signal Processing ({ICASSP})},
    year         = {2022},
}

@inproceedings{Oota24-SpeechLMs,
    author       = {Subba Reddy Oota and Emin {\c{C}}elik and Fatma Deniz and Mariya Toneva},
    title        = {Speech language models lack important brain-relevant semantics},
    booktitle    = {Proceedings of the 62nd Annual Meeting of the Association for Computational Linguistics ({ACL})},
    pages        = {8503--8528},
    year         = {2024},
}

@article{Wang25-LingNeuralDecoding,
    author       = {Yu Wang and Heyang Liu and Yuhao Wang and Chuan Xuan and Yixuan Hou and Sheng Feng and Hongcheng Liu and Yusheng Liao and Yanfeng Wang},
    title        = {Progress, challenges and future of linguistic neural decoding with deep learning},
    journal      = {Communications Biology},
    volume       = {8},
    pages        = {1350},
    year         = {2025},
}

@article{Han25-DIVER,
    author       = {Danny Dongyeop Han and Yonghyeon Gwon and Ahhyun Lucy Lee and Taeyang Lee and Seong Jin Lee and Jubin Choi and Sebin Lee and Jihyun Bang and Seungju Lee and David Keetae Park and Shinjae Yoo and Chun Kee Chung and Jiook Cha},
    title        = {DIVER-1: Scaling Intracranial EEG Foundation Models for Transferable Representations},
    journal      = {arXiv preprint},
    volume       = {arXiv:2512.19097},
    year         = {2025},
}

@article{zada25-podcast,
    author       = {Zaid Zada and Samuel A. Nastase and Bobbi Aubrey and Itamar Jalon and Sebastian Michelmann and Haocheng Wang and Liat Hasenfratz and Werner Doyle and Daniel Friedman and Patricia Dugan and Lucia Melloni and Sasha Devore and Adeen Flinker and Orrin Devinsky and Ariel Goldstein and Uri Hasson},
    title        = {The ``Podcast'' {ECoG} dataset for modeling neural activity during natural language comprehension},
    journal      = {Scientific Data},
    year         = {2025},
    volume       = {12},
    number       = {1},
    pages        = {1135},
}

@inproceedings{wang23-brainbert,
    author       = {Christopher Wang and Vighnesh Subramaniam and Adam Uri Yaari and Gabriel Kreiman and Boris Katz and Ignacio Cases and Andrei Barbu},
    title        = {BrainBERT: Self-supervised representation learning for intracranial recordings},
    booktitle    = {Proceedings of the International Conference on Learning Representations ({ICLR})},
    year         = {2023},
}

@inproceedings{chau25-popt,
    author       = {Geeling Chau and Christopher Wang and Sabera Talukder and Vighnesh Subramaniam and Saraswati Soedarmadji and Yisong Yue and Boris Katz and Andrei Barbu},
    title        = {Population Transformer: Learning Population-level Representations of Neural Activity},
    booktitle    = {Proceedings of the International Conference on Learning Representations ({ICLR})},
    year         = {2025},
}

@article{goldstein25-unified,
    author       = {Ariel Goldstein and Haocheng Wang and Leonard Niekerken and Mariano Schain and Zaid Zada and Bobbi Aubrey and Tom Sheffer and Samuel A. Nastase and Harshvardhan Gazula and Aditi Singh and Aditi Rao and Gina Choe and Catherine Kim and Werner Doyle and Daniel Friedman and Sasha Devore and Patricia Dugan and Avinatan Hassidim and Michael Brenner and Yossi Matias and Orrin Devinsky and Adeen Flinker and Uri Hasson},
    title        = {A unified acoustic-to-speech-to-language embedding space captures the neural basis of natural language processing in everyday conversations},
    journal      = {Nature Human Behaviour},
    volume       = {9},
    number       = {5},
    pages        = {1041--1055},
    year         = {2025},
    doi          = {10.1038/s41562-025-02105-9},
}

@article{jain24-computational,
    author       = {Shailee Jain and Vy A. Vo and Leila Wehbe and Alexander G. Huth},
    title        = {Computational language modeling and the promise of in silico experimentation},
    journal      = {Neurobiology of Language},
    volume       = {5},
    number       = {1},
    pages        = {80--106},
    year         = {2024},
    publisher    = {MIT Press},
}

@inproceedings{antonello23-scaling,
    author       = {Richard J. Antonello and Aditya Vaidya and Alexander G. Huth},
    title        = {Scaling laws for language encoding models in {fMRI}},
    booktitle    = {Proceedings of the Conference on Neural Information Processing Systems ({NeurIPS})},
    pages        = {21895--21907},
    year         = {2023},
    publisher    = {Curran Associates, Inc.},
}

@inproceedings{li22-blip,
  title={{BLIP}: Bootstrapping language-image pre-training for unified vision-language understanding and generation},
  author={Li, Junnan and Li, Dongxu and Xiong, Caiming and Hoi, Steven},
  booktitle={International Conference on Machine Learning (ICML)},
  pages={12888--12900},
  year={2022},
}

@inproceedings{elizalde23-clap,
  title={{CLAP} learning audio concepts from natural language supervision},
  author={Elizalde, Benjamin and Deshmukh, Soham and Al Ismail, Mahmoud and Wang, Huaming},
  booktitle={Proceedings of the IEEE International Conference on Acoustics, Speech and Signal Processing (ICASSP)},
  year={2023},
}

\end{document}